\documentclass[aps,pra,superscriptaddress,twocolumn]{revtex4-1}

\usepackage{graphicx}
\usepackage{amssymb,amsfonts,amsmath,amsbsy,wasysym,bbm}
\usepackage{epstopdf,epsfig}
\usepackage[english]{babel}
\usepackage{color}
\usepackage{enumerate}
\usepackage{times}

\definecolor{dred}{rgb}{.8,0.2,.2}
\definecolor{dgreen}{rgb}{.2,0.5,.2}
\definecolor{ddred}{rgb}{.8,0.5,.5}
\definecolor{dblue}{rgb}{.2,0.2,.8}

\newcommand{\half}{\mbox{$\textstyle \frac{1}{2}$}}

\newcommand{\ket}[1]{\vert{#1}\rangle}
\newcommand{\bra}[1]{\langle{#1}\vert}
\newcommand{\an}[1]{{\hat{#1}}}
\newcommand{\cre}[1]{{\hat{#1}}^\dagger}  

\newcommand{\tr}{\mathrm{tr}}
\newcommand{\av}[1]{\langle #1\rangle}
\newcommand{\abs}[1]{\lvert #1\rvert}
\newcommand{\id}{\mathbbm{1}}

\newcommand{\sigx}{\an{\sigma}_{x}}
\newcommand{\sigy}{\an{\sigma}_{y}}
\newcommand{\sigz}{\an{\sigma}_{z}}
\newcommand{\sigp}{\an{\sigma}_+}
\newcommand{\sigm}{\an{\sigma}_-}

\newcommand{\ee}{\mathrm{e}}
\newcommand{\ii}{\mathrm{i}}
\newcommand{\dd}{\mathrm{d}}

\newcommand{\rr}{\mathbf{r}}
\newcommand{\LL}{\mathbf{L}}

\newcommand{\qq}{\mathbf{q}}
\newcommand{\pp}{\mathbf{p}}
\newcommand{\vv}{\mathbf{v}}

\newcommand{\Ac}{\mathcal{A}}
\newcommand{\Bc}{\mathcal{B}}

\newcommand{\Vc}{\mathcal{V}}
\newcommand{\Jc}{\mathcal{J}}

\newcommand{\tabr}[1]{Table~\ref{#1}}
\newcommand{\eqr}[1]{Eq.~(\ref{#1})}
\newcommand{\fir}[1]{Fig.~\ref{#1}}
\newcommand{\secr}[1]{Sec.~\ref{#1}}
\newcommand{\apr}[1]{App.~\ref{#1}}

\begin{document}

\title{Non-destructive selective probing of phononic excitations in a cold Bose gas using impurities}

\author{D. Hangleiter}

\affiliation{Clarendon Laboratory, University of Oxford, Parks Road, Oxford OX1 3PU, UK}
\affiliation{Fakult\"at f\"ur Physik, Ludwig-Maximilians-Universit\"at, 80799 M\"unchen, Germany}

\author{M. T. Mitchison}

\affiliation{Quantum Optics and Laser Science Group, Blackett Laboratory, Imperial College London,
London SW7 2BW, UK}

\affiliation{Clarendon Laboratory, University of Oxford, Parks Road, Oxford OX1 3PU, UK}

\author{T. H. Johnson}
\email{tomi.johnson@physics.ox.ac.uk}

\affiliation{Centre for Quantum Technologies, National University of Singapore, 3 Science Drive 2, 117543 Singapore, Singapore}

\affiliation{Clarendon Laboratory, University of Oxford, Parks Road, Oxford OX1 3PU, UK}

\affiliation{Keble College, University of Oxford, Parks Road, Oxford OX1 3PG, UK}

\author{M. Bruderer}
\affiliation{Institut f\"ur Theoretische Physik, Albert-Einstein Allee 11, Universit\"at Ulm, 89069 Ulm, Germany}

\author{M. B. Plenio}
\affiliation{Institut f\"ur Theoretische Physik, Albert-Einstein Allee 11, Universit\"at Ulm, 89069 Ulm, Germany}

\affiliation{Quantum Optics and Laser Science Group, Blackett Laboratory, Imperial College London,
London SW7 2BW, UK}

\author{D. Jaksch}
\affiliation{Clarendon Laboratory, University of Oxford, Parks Road, Oxford OX1 3PU, UK}

\affiliation{Centre for Quantum Technologies, National University of Singapore, 3 Science Drive 2, 117543 Singapore, Singapore}

\affiliation{Keble College, University of Oxford, Parks Road, Oxford OX1 3PG, UK}

\date{\today}

\begin{abstract}
We introduce a detector that selectively probes the phononic excitations of a cold Bose gas. The detector is composed of
a single impurity atom confined by a double-well potential, where the two lowest eigenstates of the impurity form an effective probe
qubit that is coupled to the phonons via density-density interactions with the bosons. The system is analogous to a two-level atom coupled to
photons of the radiation field. We demonstrate that tracking the evolution of the qubit populations allows probing
both thermal and coherent excitations in targeted phonon modes.  The targeted modes are selected in both energy and momentum by
adjusting the impurity's potential. We show how to use the detector to observe coherent density waves and to measure
temperatures of the Bose gas down to the nano-Kelvin regime.  We analyze how our scheme could be realized experimentally,
including the possibility of using an array of multiple impurities to achieve greater precision from a single experimental run.
\end{abstract}

\maketitle


\section{Introduction}

Cold atomic gases play a key role in emerging quantum technologies, from the simulation of fundamental
physics~\cite{Lewenstein-AdvPhys-2007,Lewenstein-OUP-2012,Johnson-EPJ-2014} and computation~\cite{Jaksch-PRL-1999,Mandel-Nature-2003}, to time-keeping~\cite{Bize-JPhysB-2005}.
It is therefore essential that, as well as control, we are able to accurately probe the properties of atomic gases. Minimally invasive measurement schemes are of particular interest, since it is often crucial that this probing disturbs the gas as little as possible.

With this general aim in mind, we here build on a growing body of work, largely theoretical, in which a small quantum system --- a probe --- is coupled to the system of interest and then measured in order to extract information about that system. This has previously been shown to allow the extraction of information about bandwidth and gaps in the excitation spectrum~\cite{Johnson-PRA-2011}, non-equilibrium work distributions~\cite{Dorner-PRL-2013,Mazzola-PRL-2013}, temperature~\cite{Bruderer-NJP-2006,Sabin-arxiv-2013}, non-Markovianity~\cite{Haikka-PRA-2011,Haikka-PRA-2013,Rivas-RPP-2014}, effective Hamiltonian parameters~\cite{Recati-PRL-2005}, phase transitions~\cite{Punk-PRA-2013}, the Gibbons-Hawking effect \cite{Fedichev-PRL-2003}, and the Unruh effect~\cite{Retzker-PRL-2008}, often in a cold atom setting. The recent surges in experimental control of such systems~\cite{Palzer-PRL-2009,Will-PRL-2011,Spethmann-PRL-2012, Catani-PRA-2012} bring their realization within reach.

In our case, we consider a weakly-interacting Bose gas described by its phononic excitations above a condensate. We devise a detector to probe coherent and thermal occupation of a tunable subset of these modes. Detecting coherent excitations of a variety of wavelengths is important, for instance, when Bose gases are used to simulate gravitational models~\cite{Barcelo-LRevRel-2011}, or for the study of dispersive shock waves~\cite{Dutton-Science-2001,Damski-PRA-2004,Hoefer-PRA-2006}. 
Thermal excitations store information about the temperature of a gas, and thus selective probing of these acts as a thermometer. Obtaining accurate estimates of temperature is essential in many uses of cold gases in quantum technologies.

\begin{figure}[b]
\begin{center}
\includegraphics[width=\columnwidth]{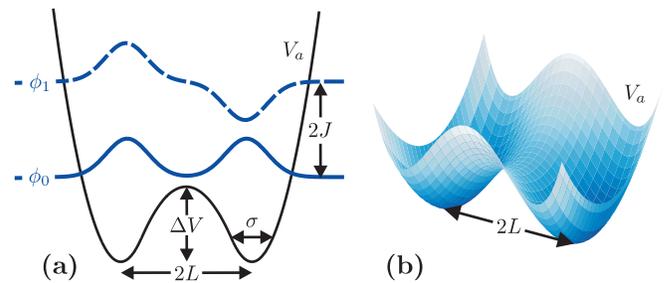}
\end{center}
\caption{(Color online) (a) The phonon detector consists of a single impurity atom in a double-well potential $V_{a}$ with well width $\sigma$ and separation $2L$, and barrier height $\Delta V$. The impurity is restricted to the two lowest states with symmetric and antisymmetric wave functions $\phi_0$ and $\phi_1$, separated by the energy splitting $2J$. This detector is immersed in a weakly-interacting Bose gas and acts as a probe. (b) In two and three dimensions, the impurity is trapped by a potential that has the same double-well shape in one direction, but is tightly confined in the remaining orthogonal directions.}

\label{setup}
\end{figure}

Standard techniques to probe Bose gases are time-of-flight (TOF), and {\it in-situ} phase-contrast~\cite{Ketterle-1999} or absorption~\cite{Armijo-PRL-2012} imaging~\cite{Bloch-RMP-2008}. From the velocity distribution measured in TOF one can infer properties of the underlying state of the Bose gas, in particular, information about both static (e.g.\ temperature) and dynamical properties (e.g.\ sound propagation)~\cite{Ketterle-1999}. However, the achievable precision in, say, a temperature measurement decreases as the nano-Kelvin regime is approached: for example, a precision of 10~\% was reported in Ref.~\cite{Leanhardt-Science-2003}. 
Moreover, if the expanding clouds are too dense, TOF imaging is no longer reliable~\cite{Meppelink-PRA-2010}. {\it In-situ} imaging, suitable also for dense clouds, is inherently limited in resolution by the wavelength of light. 
Both TOF and {\it in-situ} imaging constitute a {\it destructive} measurement on the Bose gas.

In contrast, the detector proposed here is potentially {\em non-destructive}. We analyze the potential and limitations of the detector, showing that, in principle, accurate measurements of temperatures on the order of nano-Kelvin are achievable, and that coherent density waves can be detected even if their wavelength is smaller than that of light. 

The detector, shown in \fir{setup}, comprises an impurity forming a qubit from the two lowest eigenstates of a double-well potential. The density-density interaction of the impurity and bosons translates into a multi-mode quantum Rabi type qubit-phonon interaction, capable of inducing Rabi oscillations (coherently occupied modes) and equilibration described by rate equations (thermally occupied modes). In either case the detector's evolution probes phonon modes of energy equal to the qubit energy splitting and wavelengths compatible with the distance between the two wells. The sensitivity to different energies and momenta may thus be tuned by judiciously adjusting the double-well potential. It is this flexibility in selecting energies and momenta over a range relevant to the Bose gas that is responsible for the success of the double-well impurity as a probe, in contrast, for example, to a qubit formed by the internal states of an atom in a single well.

This paper is organized as follows. First, in \secr{sec:model} we introduce the detector-phonon system and the model describing it. Second, in \secr{detector evolution} we derive the equations of motion for the detector in the presence of both thermal and coherent excitation of the phonon modes. In \secr{sec:application} we then show how tracking this evolution allows one to measure the temperature of the bath and to detect the coherent occupation of a mode. Finally, in \secr{prepmes} we discuss how the detector may be implemented experimentally, focusing on simultaneous measurements using multiple impurities, before concluding in \secr{sec:conclusion}. Details of our analysis are left to the appendices.


\section{Model}
\label{sec:model}

Our detector consists of a single impurity atom of species $\Ac$ confined by
a double-well potential $V_a$ (cf.\ \fir{setup}). We assume that the impurity
constitutes a qubit formed by the ground and the first excited states of the potential, $\ket{0}$ and $\ket{1}$ respectively, with corresponding wavefunctions $\phi_0(\mathbf{r}) = \av{\mathbf{r}\lvert 0}$ and $\phi_1(\mathbf{r})=\av{\mathbf{r}\lvert 1}$. The tunneling between the two wells results in an energy splitting of $2J$ between these energy eigenstates. The Hamiltonian of the detector reads
\begin{equation}\label{hprobe}
    \hat{H}_a = J\sigz,
\end{equation}
where the population inversion is represented by the usual Pauli matrix $\sigz = \ket{1}\bra{1} - \ket{0}\bra{0}$. We use units with $\hbar = 1$ throughout.

The dilute Bose gas is composed of bosons of species~$\Bc$ and confined by a
shallow potential $V_b$, so that the gas is practically homogeneous on the
length scale $L$ of the impurity. The bosons interact weakly with interaction strength $g
= 4\pi a_b/m_b$, where $m_b$ is the boson mass and $a_b$ is the boson-boson
$s$-wave scattering length. At low temperatures the bosons are condensed and
it is sufficient to consider the excitations, known as Bogoliubov phonons,
on top of the condensate wavefunction $\psi_0(\rr) = \sqrt{n_0}$ with number density $n_0$. In this approximation the
Hamiltonian of the Bose gas, up to a constant, is~\cite{Oosten-PRA-2001}
\begin{equation}\label{hphonon}
    \an{H}_b = \sum_{\qq\neq 0}\omega_\qq \cre{b}_\qq \an{b}_\qq\,,
\end{equation} 
where the operator $\cre{b}_\qq$ ($\an{b}_\qq$) creates (annihilates) a
Bogoliubov phonon with momentum $\qq$ and energy $\omega_\qq = \sqrt{E_\qq
(E_\qq + 2 g n_0)}$. Here, $E_\qq =
|\qq|^2/2m_b$ is the free-particle energy.

The functioning of the detector is based on collisions with the background gas, in which bosons are scattered by the impurity. At low energies, only $s$-wave scattering contributes significantly, with the momentum-independent scattering cross-section $4\pi a_{ab}^2$, where $a_{ab}$ is the $s$-wave scattering length for boson-impurity collisions \cite{landauQM}. We can therefore make the pseudo-potential approximation \cite{HuangSM}, writing an effective impurity-boson interaction potential as
\begin{equation}
\label{pseudopotential}
V(\an{\mathbf{r}}-\an{\mathbf{x}}) = \kappa \delta(\an{\mathbf{r}}-\an{\mathbf{x}}),
\end{equation}
where $\an{\mathbf{r}}$ and $\an{\mathbf{x}}$ denote the coordinates of the boson and impurity, respectively. The coupling constant $\kappa = 2\pi a_{ab}/m_b$ is chosen so that the Born approximation applied to \eqr{pseudopotential} predicts the correct scattering cross-section for collisions of low-energy bosons from the impurity. 

Within the pseudo-potential approximation, the effective interaction between detector and the Bogoliubov
phonons is described, up to a constant, by the Hamiltonian
\begin{equation*}
    \hat{H}_{ab} = \sum_{\mu,\nu}\hat{B}_{\mu\nu}\otimes \ket{\mu}\bra{\nu} \, ,
\end{equation*}
where $\mu,\nu\in\{0,1\}$ and the operator elements $\hat{B}_{\mu\nu} = \kappa \sum_{\qq\neq 0} (M_{\qq,\mu\nu} \cre{b}_\qq + \mathrm{h.c.} )$
are written in terms of matrix elements $M_{\qq,\mu\nu}$ specifying the detector-phonon coupling.
Under the same approximations as used to obtain $\an{H}_a$ and $\an{H}_b$ with the additional
assumption of a weak coupling $\kappa$ we find~\cite{Oehberg-PRA-1997,Oosten-PRA-2001}
\begin{equation*}
    M_{\qq,\mu\nu} = S_\qq \int\!\dd \rr \,\phi_\mu(\rr)\phi^*_\nu( \rr)\ee^{\ii \qq \cdot \rr}\,,
\end{equation*}
with $S_\qq = \sqrt{n_0 E_\qq/\Vc \omega_\qq}$ and $\Vc$ the volume of the Bose gas.

The pseudo-potential approximation is valid if any excitation to higher vibrational states of the impurity can be neglected, and if the initial wave-vector $\mathbf{q}$ of the scattered atom satisfies $qr_0\gg 1$, where $r_0$ is the effective range of the potential. The range $r_0$ is expected to be on the order of the scattering length $a_{ab}$ \cite{Leggett2001RMP}. Note that, due to the presence of the impurity trapping potential, the length $a_{ab}$ may differ from the bare impurity-boson scattering length~\cite{Naidon-PRA-2006,Naidon-NJP-2007}.

\subsection{Symmetric double well in the harmonic approximation}

The properties of the detector are determined by the qubit wavefunctions $\phi_\mu (\rr)$, which are in turn determined by the shape of the double-well potential $V_{a}(\rr)$. From here on we consider tractable types of potential $V_{a}(\rr)$ and approximate wavefunctions $\phi_\mu (\rr)$ that enable us to analytically study the dependence of the detector on its potential. Note that the accuracy of the detector itself does not rely on the accuracy of this assumption. 

\begin{figure}[t]
\begin{center}
\includegraphics[width=\columnwidth]{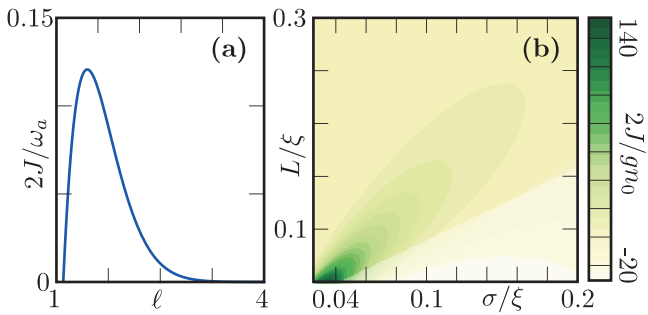}
\end{center}
\caption{(Color online) {(a)} Dependence of the energy splitting $2J/\omega_a$ according to \eqr{J} as a function of (a) the relative well separation $\ell$, and {(b)} $L$ and $\sigma$ in terms of the natural Bose gas length and energy scales, the healing length $\xi = 1/\sqrt{m_b g n_0}$ and chemical potential $g n_0$. The shaded region in the bottom right corner is inaccessible since here $L < \sigma$.
The other parameters for both plots are $D=1$, $m_a = 0.5 \, m_b$ and $\Delta V/\omega_a = D$.}
\label{fig: J}
\end{figure}

Explicitly, we consider a deep symmetric potential with a local maximum at $\rr = 0$ and two minima at $\rr = \pm \LL$ such that the qubit states are approximately described by symmetric and antisymmetric superpositions of Gaussian wavefunctions $\phi (\rr) =
(\pi\sigma^2)^{-D/4} \exp(-r^2/ 2\sigma^2)$ of width $\sigma = 1 / \sqrt{ m_a \omega_a}$ centered at the respective double well minima $\rr = \pm \LL$. Here $m_a$ is the mass of the impurity. In this case the energy splitting $2J$ between the qubit states obeys~\cite{Dounas-Frazer-thesis}
\begin{equation}\label{J}
\frac{2J}{\omega_a} =  \left[\left(\ell^2 - \frac{1}{2} \right) - D + 1 - \frac{2\Delta V }{\omega_a} \right] \ee^{-\ell^2}  \,,
\end{equation}
where $\ell = L/\sigma$. This is shown in Figs.~\ref{fig: J}(a) and (b) with respect to the detector and Bose gas length and energy scales, respectively. 

The excitations to higher vibrational levels of the double well potential can be neglected in the regime where $J/\omega_a\ll 1$. Since $J/\omega_a$ decays exponentially with $\ell$ this qubit regime can be readily achieved.

For these wavefunctions, the interaction matrix elements become 
\begin{equation}\label{simpel}
\begin{split}
    M_{\qq,01} &=  M_{\qq,10} = -\ii S_\qq\sin(\qq \cdot \LL)\,\ee^{-\sigma^2 q^2/4} \,, \\
    M_{\qq,00} &= S_\qq\big[\cos(\qq \cdot \LL) + \ee^{-\ell^2}\big]\,\ee^{-\sigma^2 q^2/4} \,, \\
    M_{\qq,11} &= S_\qq\big[\cos(\qq \cdot \LL) - \ee^{-\ell^2}\big]\,\ee^{-\sigma^2 q^2/4} \,.
\end{split}
\end{equation}
The difference between the diagonal matrix elements is negligible when $\ell \gtrsim 1$ since
$M_{\qq,00} - M_{\qq,11} \sim \ee^{-\ell^2}$. We may therefore re-express
the interaction Hamiltonian in this regime as
\begin{equation}\label{rehab}
	\hat{H}_{ab} = \hat{B}_{10} \otimes \sigx + \hat{B}_{00} \otimes \id \, .
\end{equation}
The full Hamiltonian described by Eqs. \eqref{hprobe}, \eqref{hphonon} and \eqref{rehab} represents a multi-mode quantum Rabi model, which forms the basis for all of the following analysis. The operator elements are $\hat{B}_{10} = \kappa(\an{n}_L - \an{n}_R)$ and $\hat{B}_{00} = \kappa(\an{n}_L + \an{n}_R)$,
where $\an{n}_{L} = \int \dd \rr | \phi (\rr + \LL) |^2 \an{n} (\rr) $ is the density deviation $\an{n} (\rr) = \sum_{\qq \neq 0} S_\qq (\ee^{\ii \qq \cdot \rr} \cre{b}_\qq + \mathrm{h.c.})$ of the Bose gas from $n_0$, averaged over the left well, and $\an{n}_{R}$ is similarly defined. The Hamiltonian in \eqr{rehab} describes the coherent driving of the qubit by the difference between the boson densities at the two wells. 

As we explain in detail in subsequent sections, the detector is sensitive to phonon modes whose energy and wavelength approximately coincide with the corresponding energy and length scales of the detector, namely $2J$, $L$ and $\sigma$. In \apr{app:experimental realization} we describe a possible experimental realization of our detector consisting of an array of double-well potentials created by an optical
lattice~\cite{Sebby-Strabley-PRL-2007,Foelling-Nature-2007}. Alternatively, optical potentials that are tailored through the modulation of an optical dipole trap may be used~\cite{Zimmermann-NJP-2011}. We demonstrate that a broad range of frequencies and momenta may be accessed by modifying the impurity trapping potential $V_a$ within the limits of what is possible using current technology. Importantly, we find that the double-well impurity detector is sensitive to the normal energy and momentum scales corresponding to a typical Bose gas, but can also probe a much broader spectrum of frequencies ranging from several Hertz up to values on the order of $100$ kHz. In Bose gases, the lower limit corresponds to nano-Kelvin temperatures, while the upper limit corresponds to phonon modes with sub-optical wavelengths. Our proposed detector therefore could enable the measurement of very low temperatures (at the low frequency extreme) or short-distance density variations (at the high frequency extreme), which are difficult to resolve by alternative means. In the examples which follow we mainly focus on these extreme cases, since they most aptly illustrate the capabilities and limitations of the detector. We also use that coherence times of the detector of more than 10 ms are readily achieved~\cite{Anderlini-Nature-2007, Foelling-Nature-2007}.


\section{Evolution of the Detector}
\label{detector evolution}

In order for the detector to probe the Bose gas it must interact with it for a time $\tau$ and then be measured. 
The operation of the detector is determined by how the reduced density operator of the impurity $\hat{\rho}_a(t)=\tr_b\{\hat{\rho}(t)\}$ evolves, with
\begin{equation}\label{redeq}
    \frac{\partial}{\partial t}\hat{\rho}_a(t) = -\ii[\hat{H}_a,\hat{\rho}_a(t)] -\ii\,\tr_b\{[\hat{H}_{ab},\hat{\rho}(t)]\}\,,
\end{equation}
the equation governing this evolution, $\hat{\rho}(t)$ the density operator of the total system and $\tr_b\{\,\cdot\,\}$ the partial trace over the states of the Bose
gas. In the following we solve \eqr{redeq} in different approximations
to determine the effect of different states of the Bose gas on the detector.

\subsection{Damping by the thermal phonon bath}
\label{damping}

To start with, we analyze how the double-well qubit responds to thermal fluctuations of the Bose gas. The appropriate tool for describing the evolution of $\hat{\rho}_a$ is a quantum master
equation in Lindblad form, derived under the standard Born-Markov and rotating wave approximations \cite{breuer2007book}. This equation takes the form
\begin{equation}\label{QME}
\frac{\partial\hat{\rho}_a (t)}{\partial t} = -\ii[\hat{H}_a,\hat{\rho}_a] +  \sum_{s=\pm} k_{s}\Big(\an{\sigma}_s\hat{\rho}_a(t)\cre{\sigma}_s \\
-\frac{1}{2}\{\cre{\sigma}_s\an{\sigma}_s,\hat{\rho}_a(t)\}\Big),
\end{equation}
where $\an{\sigma}_{\pm} = \half (\sigx \pm \ii\sigy)$. In writing \eqr{QME}, the small renormalization (Lamb shift) of the detector energy levels due to the coupling with the gas has been neglected. The final term in the equation describes stochastic transitions between populations of the detector, occurring at the rates
\begin{equation}
\begin{split}
	k_- &= 2\pi [n(2J)+1] \Jc_D(2J)\,, \\
	k_+ &= 2\pi n(2J)\Jc_D(2J)\, ,
\end{split}	
\label{dampingrates}
\end{equation}
where $n(\omega) = [\exp (\omega/ k_B T) - 1]^{-1}$ is the average number of phononic excitations of energy $\omega$. The spectral density is given by
\begin{equation}\label{sdens}
	\Jc_D(\omega) = \kappa^2\sum_{\qq\neq 0}|M_{\qq,10}|^2 \delta(\omega-\omega_\qq).
\end{equation}

The evolution according to the master equation \eqref{QME} has the favorable property
that populations and coherences of $\hat{\rho}_a$ evolve independently, as
\begin{equation}\label{eq:rho11}
\begin{split}
	\rho_{11}(t) &= \left[\rho_{11}(0)-\frac{k_+}{2\bar{k}}\right]\ee^{-2\bar{k}t} + \frac{k_+}{2\bar{k}}\,, \\
	\rho_{10}(t) &= \rho_{10}(0)\ee^{-\ii 2J t - \bar{k}t}\,,	
\end{split}
\end{equation}
where $\bar{k} = \frac{1}{2}(k_+ + k_-)$ is the average rate, while $\rho_{\mu\nu}(t) = \tr_a \{ \an{\rho}_a (t) \ket{\mu}\bra{\nu} \}$, with $\mu,\nu \in \{ 0 , 1 \}$, are the reduced density matrix elements. We see that the coherences $\rho_{10}$ of the double-well qubit
decay exponentially due to the interaction with the thermal phonon bath. Accordingly, in the long-time limit, the qubit evolves towards
equilibrium with the phonon bath, while the coupling strength remains unchanged. That is, the qubit evolves towards a mixed state with asymptotic populations $\rho_{00} = k_{-}/(k_{+} + k_{-})$ and $\rho_{11} = k_{+}/(k_{+} + k_{-})$ and vanishing coherences $\rho_{10}$, which corresponds to a thermal state of temperature $T$.

To give some insight into the characteristics of the equilibriation process, we evaluate the spectral density \eqref{sdens} in the regime of large impurity widths $\sigma \gg \xi$, where $\xi = 1/\sqrt{m_b g n_0}$ is the healing length of the Bose gas. In this regime we can replace the Bogoliubov dispersion relation with its low-frequency approximation $\omega_\qq = c |\qq|$, with $c = \sqrt{gn_0/m_b}$ the speed of sound. In addition, we take the usual continuum limit $\Vc^{-1} \sum_{\qq\neq 0} \rightarrow (2\pi)^{-D} \int \dd \qq$. In terms of the characteristic frequencies $\omega_L = c/L$ and $\omega_\sigma = c/\sigma$, we then obtain
\begin{equation}\label{speceq}
\begin{split}
	\Jc_{\rm 1}(\omega) &= \frac{\kappa^2}{2\pi g c}\omega \sin^2(\omega/\omega_L) \ee^{-\frac{1}{2}(\omega/\omega_\sigma)^2}\,, \\[2mm]
	\Jc_{\rm 2}(\omega) &= \frac{\kappa^2}{8\pi g c^2}\omega^2 \left[1-\Bc_0(2\omega/\omega_L)\right]\ee^{-\frac{1}{2}(\omega/\omega_\sigma)^2}\,, \\[2mm]
	\Jc_{\rm 3}(\omega) &= \frac{\kappa^2}{8\pi^2 g c^3} \omega^3 \left[1-\mathrm{sinc}(2\omega/\omega_L)\right]\ee^{-\frac{1}{2}(\omega/\omega_\sigma)^2}\,.
\end{split}
\end{equation}  
Here, $\Bc_0(x)$ is a Bessel function of the first kind. 
The detailed low-frequency shape of the spectral density depends on the frequency $\omega_L$, which is set by the well
distance, whereas the high-frequency cut-off is determined by~$\omega_\sigma$.

\begin{figure}[t]
\begin{center}
\includegraphics[width=\columnwidth]{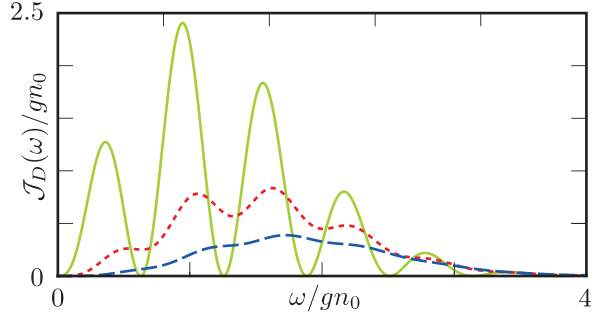}
 \end{center}
\caption{(Color online) Spectral densities $\Jc_{D}(\omega)$ in the linear regime [\eqr{speceq}] for $D = 1$ (green solid), 2 (red dotted), 3 (blue dashed). The detector parameters used are $\sigma = \xi$, $L = 5\, \xi$ and $\Delta V = \omega_a$, the detector-Bose gas coupling is $\kappa = 5\, g$, and the Bose gas density $n_0 \xi^D = 1$. }
\label{spectral}
\end{figure}

Figure \ref{spectral} shows the spectral densities for different dimensions, evaluated in the low-frequency approximation. For
the one-dimensional Bose gas, the spectral density exhibits several maxima and vanishes periodically between these.
The qubit is thus completely decoupled from the Bose gas for specific values of the energy splitting $2J$. The
origin of this structure is energy and momentum conservation, which in one dimension depends strictly on the matching
between energy $\omega_q$ and momentum $q$ of the phonons and the energy splitting $2J$ and size $L$ of the detector.
For a two- and three-dimensional Bose gas, the structure of the spectral density is significantly less pronounced. In
this case, the momentum $\qq$ projected onto the direction of the detector $\LL$ is the relevant conserved quantity and
phonons impinging on the detector from different angles can always fulfill energy and momentum conservation. For use outside the sound-like regime, we have evaluated the spectral
density using the full Bogoliubov dispersion relation in \apr{app:spectral density}. 

To conclude this section, we briefly discuss the range of validity of the master equation. The Born-Markov approximation is
valid so long as the dissipative evolution induced by the bath occurs slowly compared to the thermal correlation time of the
bath, i.e.\  $k_{\pm} \ll k_B T$. The spontaneous emission rate $k_{-}$ must also be much slower than the vacuum correlation
time $\tau_v$, corresponding to the inverse of the bandwidth of Bogoliubov frequencies that interact appreciably with the impurity.
The vacuum correlation time can be estimated as $\tau_v \approx \sigma/c$ when $\sigma \gtrsim \xi$, or
$\tau_v \approx m_b \sigma^2$ when $\sigma \lesssim \xi$. Finally, the rotating wave approximation is applicable only
if the coherent evolution is significantly faster than the dissipative evolution, i.e.\ $J\gg k_{\pm}$. In three dimensions,
typical values of the relevant parameters are $\tau_v \sim 150 \, \mu$s and $1/k_BT \sim  10 \, \text{ms}$ at $T = 1 \, \text{nK}$
for an impurity size of $\sigma = 100 \, \text{nm}$. Hence, for Bose gas lifetimes on the order of $1$s all requirements can
be readily fulfilled.

\subsection{Driving by coherent phonons}
\label{coherent driving}

Next, we assume that the detector and the Bose gas, described by
$\hat{\rho}_b(t)$, are not correlated so that $\hat{\rho}(t)
=\hat{\rho}_a(t)\otimes\hat{\rho}_b(t)$ at all times $t$. In this regime
\eqr{redeq} simplifies to
\begin{equation}\label{heff}
    \frac{\partial \hat{\rho}_a }{\partial t} (t)= -\ii[\hat{H}_a + \hat{H}_{m\!f}(t),\hat{\rho}_a(t)]\,,
\end{equation}
where we have introduced the mean-field Hamiltonian $\hat{H}_{m\!f}(t) =
\tr_b\{\hat{\rho}_b(t)\hat{H}_{ab}\}$. 

The mean-field approximation does not account for quantum fluctuations, but
it is very suitable for describing experiments with the phonon modes in
coherent states. This situation occurs naturally when a classical potential
acts on the Bose gas, e.g., a laser beam~\cite{Andrews-PRL-1997} or the density of an impurity \cite{Johnson-EPL-2012}. More
precisely, for a Bose gas prepared in the state
$\hat{\rho}_b(t) = \ket{\beta_\pp(t)}\bra{\beta_\pp(t)}$ with a single phonon mode of frequency $\omega_\pp$ occupied coherently
$\an{b}_\qq\ket{\beta_\pp} = \delta_{\pp\qq} \beta_\pp \ket{\beta_\pp}$, we obtain the
mean-field Hamiltonian
\begin{equation*}
    \hat{H}_{m\!f}(t) =   \Omega_\pp \cos(\omega_\pp t - \theta_\pp ) \sigx \, ,
\end{equation*}
which recovers the Hamiltonian of the classical Rabi model with resonant Rabi frequency $\Omega_\pp = 2 \kappa | M_{\pp,10}^* \beta_\pp |$ and initial phase $\theta_\pp = \arg ( \beta_\pp M_{\pp,10}^*)$. Making the rotating wave approximation, valid near resonance $ \Omega_\pp , |\delta_\pp | \ll \omega_\pp + 2J$, with $\delta_\pp = \omega_\pp - 2J$, we find coherent Rabi oscillations at frequency $\tilde{\Omega}_\pp = (\Omega_\pp^2 + \delta_\pp^2)^{1/2}$. Specifically, the expected population of the ground state $\rho_{00} (t)$, given that the detector  started in this state $\rho_{00} (0) = 1$, is found to be
\begin{align}\label{pdynamicscoherent}
 \rho_{00} (t) =  \frac{ \delta_\pp^2 }{ \tilde{ \Omega }_\pp^2}  + \frac{\Omega_\pp^2 }{ \tilde{ \Omega }_\pp ^2 } \cos^2 \left( \frac{\tilde{\Omega}_\pp t}{2} \right)  \, .
\end{align}

\begin{figure}[t]
\begin{center}
\includegraphics[width=\columnwidth]{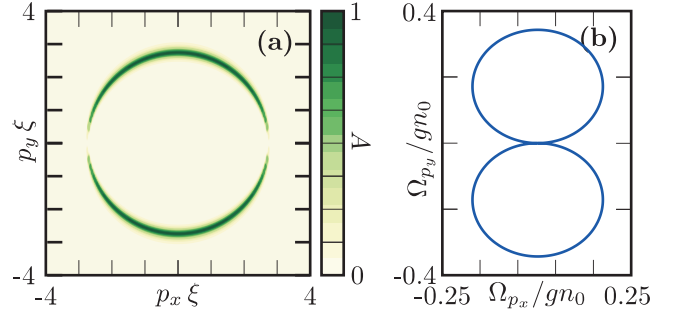}
\end{center}
\caption{(Color online) The detector in a two-dimensional setup with $m_a =0.5 \, m_b$, $\Delta V = \omega_a$, $\sigma = 0.1 / \abs{\pp}$, and $\LL  = 0.27 \, \pp/ \abs{\pp}^2$ is tuned close to resonance with a mode with the wavevector $\pp = 3 \, \hat{\mathbf{y}} / \xi$ along the y-axis. 
(a) The amplitude $A = \Omega_\pp^2/\tilde{\Omega}_\pp^2$ of coherent oscillations exhibits a clear dependence on both magnitude and direction of $\pp = (p_x, p_y)$, i.e., the direction and energy of the density waves. {(b)} Angular dependence of the Rabi frequency $\Omega_\pp$ on the direction of $\pp$. The other parameters (for both plots) are $\kappa =10 \, g$, $|\beta_\pp| = 5$, $n_0 \xi^2 = 0.5$, and $\Vc = (100 \, \xi)^2$.}
\label{oscillation amplitude}
\end{figure}

The amplitude of oscillations in the population $\Omega_\pp^2/\tilde{\Omega}_\pp^2$ is near maximum when
the detuning is small compared to the resonant Rabi frequency, i.e.~$\delta_\pp/\Omega_\pp\ll 1$.
On the other hand, the amplitude is near zero in the off-resonant case $\delta_\pp/\Omega_\pp\gg 1$.
Conservation of momentum and energy leads to a non-trivial angular dependence of the Rabi frequency
$\Omega_\pp = \Omega_\pp(\pp,\LL)$. Hence the detector is not only sensitive to the energy of the
incoming waves but also their direction. A specific example of the dependence of the oscillation amplitude
and frequency on $\pp$ is shown in Figs.~\ref{oscillation amplitude}(a) and (b). 

For the more realistic case that coherent occupation occurs in Bose gas modes with a width $\gamma$ of frequencies around $\omega_\pp$, rather than a single mode, the impurity dynamics is given by a coherent superposition of Rabi oscillations with frequency spread $\gamma$. These oscillations will go out of phase after a time $1/\gamma$ and thus will be visible if $\gamma \lesssim \tilde{\Omega}_\pp$. Assuming this to be true we hereon discuss only the single mode case.

Note that \eqr{heff} is complemented by an equivalent expression with the roles of the detector and the Bose gas interchanged,
which leads to higher-order corrections in the detector evolution. However, we assume that the evolution of the initial state
of the Bose gas $\hat{\rho}_b(0)$ according to $\partial \hat{\rho}_b (t) / \partial t = - \ii [\hat{H}_b ,\hat{\rho}_b (t)]$
dominates over the back-action of the detector. In App.~\ref{app:back-action} we show that this assumption is justified in the regime where $\Omega_\pp/\omega_\pp \ll |\beta_\pp|^2$. This places a practical limit on the amplitude of coherent density oscillations that can be measured with our detector. Specifically, we must have $\Omega_{\pp} \ll \omega_{\pp}$ in order for the rotating wave approximation to hold, therefore the back-action of the detector on the gas can be neglected if $|\beta_{\pp}|^2 \gtrsim 1$.


\subsection{Coherent driving in the presence of incoherent damping}

Naturally, some thermal excitations will always be present in the Bose gas and they cannot necessarily be ignored,
as we have assumed in \secr{coherent driving}. The full equation of motion for $\hat{\rho}_a$, including both coherent
driving and dissipative effects, is analogous to the optical Bloch equations describing a laser-driven two-level atom
damped by the free radiation field (at finite temperature), but in the context of cold atoms.

Following this analogy we may express the state of the detector in terms of the Bloch vector $\vv$ whose components
are given by $v_i = \tr_a\{\an{\rho}_a\an{\sigma}_i\}$ with $i = x,y,z$. The resulting Bloch equations in vector form read
\begin{equation}\label{bloch}
	\frac{\dd \vv}{\dd t} = \mathbf{b}\times\vv - \bar{k}\mathbf{c}\,,
\end{equation}
where $\mathbf{b} = (\Omega_{\qq},0,\delta_{\qq})^{T}$ acts as an effective magnetic field and
$\mathbf{c} = (v_x,v_y,2v_z - 2v_0)^{T}$, with $v_0 = (k_{+}-k_{-})/(k_{+}+k_{-})$
the long-time asymptotic population inversion. In the fully coherent case with $\bar{k} = 0$
the solution of the Bloch equations~\eqref{bloch} reduces to the Rabi oscillations described
by Eq.~\eqref{pdynamicscoherent}. The general analytical solution of Eq.~\eqref{bloch} is
presented in \apr{app: diagonalization}.

It follows from this solution that the double-well qubit in the presence of both coherent near-resonant driving and damping acts very similarly
to a classical receiver for electromagnetic waves, i.e.~a damped RLC circuit, that is tuned in resonance with a particular
frequency.  As such, a key quantity in determining whether there will be visible Rabi oscillations is the Q-factor
$Q = \Omega_\pp/2 \bar{k}$ relating the resonant Rabi frequency to the thermal decay.  In the sound-like regime $\sigma \gg \xi$, the $Q$-factor
\begin{equation*}
Q = \left( \frac{2 |\beta_\pp |}{2 n(2J)+1} \right) \left( \frac{\kappa S_\pp}{2\pi \Jc_D(2J)} \right) \sin(\pp \cdot \LL) \, \ee^{-\sigma^2 p^2/4} \, , 
\end{equation*}
is significantly increased when the spectral density $\Jc_D(2J)$ is near one of its nodes. As a result, counter-intuitively, we find that coherent oscillations
are often clearest when the qubit energy $2J$ is detuned away from $\omega_\pp$ towards smaller $\Jc_D(2J)$, even though, as shown
in \secr{coherent driving}, this reduces the maximum oscillation amplitude $\Omega^2_\pp / \tilde{ \Omega }^2_\pp  $. 

The other effect of damping is that the full width half maximum of the amplitude as a function of frequency gains a contribution $\sim \bar{k} = \Omega_\pp / 2 Q$ due to lifetime broadening, in addition to the usual $2 \Omega_\pp$. A low $Q$ transition is thus slightly easier to drive away from resonance.


\section{Application}
\label{sec:application}

In \secr{detector evolution} we saw that a double-well qubit evolves in response to phononic excitations in the Bose gas,
both thermal and coherent, of energy close to $2J$, which can therefore be tuned to focus on modes of interest. This
allows the measurement of the average number of coherently and thermally excited phonons present in these modes.

We focus on two examples of how our detector may be used to probe the Bose gas non-destructively. The first is the accurate measurement of the average occupation of
phonon modes of energy $2J$, from which the temperature of the Bose gas can be inferred. The second is the detection
of coherent density waves in the Bose gas. In both cases we show how the parameters of the detector can be chosen
to obtain reliable measurement results.

\subsection{Measuring the Bose gas temperature}
\label{temperature detection}

\begin{table*}[t]
  \centering
  \begin{tabular}{c|ccc|ccc|cc||c|c}
$D$ &  $k_B T_0 / g n_0$ & $k_B T_1 / g n_0$ & $n_0 \xi^D$ & $\sigma/\xi$ & $L/\xi$ & $ J/ gn_0$ & $\bar{k}(T_0) / gn_0$ & $ \bar{k}(T_1) / gn_0$ & $(\Delta T_0/T_0)_{\text{eq.}}$ & $(\Delta T_0/T_0)_{\text{non-eq.}}$ \\ \hline \hline
    1 & 0.01 & 1 & 1 & 0.5 & 1.5 & $3.21 \cdot 10^{-3}$ & $2.16 \cdot 10^{-4}$ & $ 2.08 \cdot 10^{-2}$ & 4.64 \% & 1.67 \%\\
2 & 0.5 & 10 & 5 & 0.1 & 0.27 & $1.22 \cdot 10^{-1}$ & $5.74 \cdot 10^{-3}$ & $ 1.12 \cdot 10^{-1}$ & 5.96 \% & 0.86 \%\\
3 & 5 & 100 & 20 & 0.01 & 0.03 &  $6.17 \cdot 10^{-1}$ & $2.07 \cdot 10^{-3}$ & $ 4.12 \cdot 10^{-2}$ & 11.5 \% & 0.59 \%  
  \end{tabular}
\caption{Parameters and relative uncertainties $(\Delta T_0/T_0)_{\text{eq.}}$ for the equilibrium method according to \eqr{eq:error eq}, and $(\Delta T_0/T_0)_{\text{non-eq.}}$ for the non-equilibrium method as obtained in a simulation of the shot noise for the temperature measurement under ideal conditions (see main text and \fir{fig:incoherent examples}), where $M=5000$ measurements are performed to estimate the population $\rho_{11}(t)$ for each $t$ and $t= \infty$ for the equilibrium method. For all cases we set $\Vc = (100~\xi)^D$, $ \kappa = 15~g$, $\Delta V = D \omega_a$ and $m_a = 0.5 \, m_b$.}
\label{tab:incoherent examples}
\end{table*}


\begin{figure}[b]
\begin{center}
\includegraphics[width=\columnwidth]{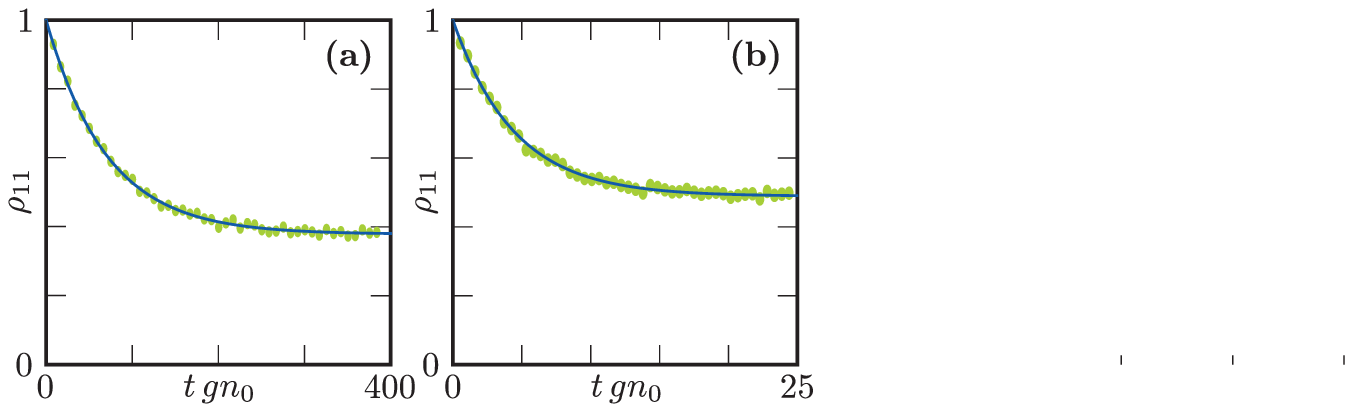}
\end{center}
\caption{(Color online) Simulation of the temperature measurement in two dimensions under ideal conditions using the values presented in \tabr{tab:incoherent examples}.
The detector is initialized in its excited state $\ket{1}$ and evolves according to \eqr{eq:rho11}. The fictional measurement
points (green) are distributed around the expectation value $\rho_{11}(t)$ according to a Bernoulli distribution obtained from
$5000$ measurements at each of the $50$ points. The corresponding plots show the decay of the population $\rho_{11}(t)$ for temperatures
{(a)} $T_0 = 0.5~gn_0/k_B$ and (b) $T_1 = 10~gn_0/k_B$, resulting in the markedly different decay times.}
\label{fig:incoherent examples}
\end{figure}

We first show how the detector can be used to measure the temperature $T_0$ of the Bose gas in the nano-Kelvin regime. We present two different methods, {\it (a)} an {\it equilibrium method}, and {\it (b)} a {\it non-equilibrium method}, either of which is capable of accessing low temperatures as low as $T_0 \approx 2 J/k_B$, which can be lowered into the nano-Kelvin regime. 


\paragraph{Equilibrium method.} 
A conventional thermometer relies on the measurement of the population after full equilibration with the environment. To use the double-well qubit as such a thermometer it is first allowed to come to equilibrium with the Bose gas, then the excited-state population $\rho_{11} = k_{+}/(k_{+} + k_{-})$ is measured. Knowledge of the population $\rho_{11}$ and the splitting $2J$ is therefore sufficient to determine the Bose-gas temperature $T_0$. The advantage of this method is that it does not rely on the physics specific to the system whose temperature is being measured. Specifically, the accuracy of the thermometer does not rely on the accuracy of the Bogoliubov description of the Bose gas.

A precise measurement of $T_0$ using this method requires that $n(2J,T_0) \approx 1$, or equivalently $2 J/k_B T_0 \approx 1$, so that the population $\rho_{11}(T_0)$ is sufficiently sensitive to the value of $T_0$. This means that it is possible to measure temperatures $T_0$ on the order of $2J/k_B$. More specifically, in \apr{sec:Terror} we show that the uncertainty in estimating $\rho_{11}(t)$ using $M$ measurements of the state of the qubit in the energy basis contributes an uncertainty $\Delta T_0$ when measuring $T_0$ obeying
\begin{equation}
  \frac{\Delta T_0}{T_0} = \frac{1}{\sqrt{M}} \left[ \frac{k_B T_0}{2J} \frac{(2 n(2J,T_0) + 1)(n(2J,T_0) + 1)^{1/2}}{\ee^{2J/k_B T_0} [ n(2J,T_0)] ^{3/2}} \right] \, . \label{eq:error eq}
\end{equation}
The uncertainty $\Delta T_0$ has a global minimum of value $\sim~M^{-1/2}$ at $2 T_0 \approx J/k_B$ and diverges as $T_0$ becomes very large or very small, which is consistent with the above estimate $2 J/k_B T_0 \approx 1$.


\paragraph{Non-equilibrium method.} 
The temperature is not only revealed by the equilibrium populations, but also by how the populations decay to equilibrium. Thus another approach is to initialize the double-well qubit in state $\ket{1}$, {\it observe the decay} of the population $\rho_{11}$ towards its equilibrium value and extract the temperature-dependent decay rate $\bar{k}(T)$, from which the temperature can be deduced. To avoid any sensitivity to the precise values in the impurity-Bose gas or Bose gas Hamiltonian, beyond the assumption of a continuum of weakly-interacting bosonic modes around energy $2J$, we in fact extract the decay rates $\bar{k}(T)$ for two temperatures. The first is a (high) reference temperature $T_1$ and the second is an unknown temperature $T_0 \ll T_1$, which we determine from the ratio
\begin{equation}\label{decay}
  \frac{\bar{k}(T_0)}{\bar{k}(T_1)}  = \frac{2 n(2J,T_0) +1}{2n(2J,T_1) + 1} \, ,
\end{equation}
given the values of $T_1$ and $2J$. This process of measuring temperature by observing decay rates, rather than equilibrium populations, differs fundamentally from a conventional thermometer. 

The non-equilibrium approach requires many measurements in order to obtain a sufficiently precise estimate of $\rho_{11}$ at multiple times. However, it also has the advantage that all measurements can be carried out within a time much shorter than the equilibration time. Moreover, if the temperature is very small, whence the equilibrium value of $\rho_{11}$ becomes small, the equilibrium method is very susceptible to systematic errors in the population measurement, as well as heating due to fluctuations in the double-well potential. Since larger values of $\rho_{11}$ are obtained during the non-equilibrium scheme for measurements at small times, this method is less susceptible to such errors. 

A precise measurement of $T_0$ using the non-equilibrium method requires that $2J /k_B T_0 \lesssim 1$, or $n(2J,T_0) \gtrsim 1$. Specifically, the measurement uncertainty for the value of $T_0$ using the non-equilibrium method (derived in \apr{sec:Terror}) scales approximately as
\begin{align*}
  \frac{ \Delta T_0}{T_0} \approx \frac{1}{\sqrt{M}} \left[\frac{k_B T_0}{2 J}\frac{2 n(2J,T_0) +1}{2 \sqrt{2} \ln (2) \ee^{2J /k_B T_0} n^2(2J,T_0)} \right] \,  . 
\end{align*}
In accordance with the above estimate $2J/k_B T_0 \lesssim 1$ the quantity in square brackets is on the order of unity for $2J /k_B T_0 \approx 1$, but rises quickly for smaller temperatures.

Note that the detector parameters $L$ and $\sigma$, while keeping $2J/k_B T_0$ fixed, can be tuned such that the spectral density $\Jc_D(2J)$ and the decay rates $k_{\pm}$ result in the desired temporal resolution. In particular, the incoherent evolution of the qubit can be made sufficiently slow for the Born-Markov approximation to be valid, or fast enough to be within the coherence time of the qubit.

To illustrate the measurement process for the non-equilibrium method, we have simulated the shot noise resulting from a finite number of measurements during an ideal experiment to determine $T_0 \sim 1$~nK, with a reference temperature $T_1 \sim 100$~nK. For the Bose gas and impurity parameters presented in \tabr{tab:incoherent examples} we
obtain relative uncertainties $\Delta T_0/T_0 \lesssim 1 ~\%$ after taking $M = 5000$ measurements at each of the 50 time points. An example for $D = 2$ is depicted in \fir{fig:incoherent examples}. Likewise, under the same ideal conditions and parameters, the uncertainties for the equilibrium method according to \eqr{eq:error eq} assuming full equilibration and taking $M = 5000$ measurements are $\Delta T_0/T_0 \lesssim 10~\%$. Hence, using either method, it is, in principle, possible to precisely measure temperatures in the nano-Kelvin regime. 

\subsection{Detection of coherent density waves}
\label{coherent detection}

\begin{figure}[t]
\begin{center}
\includegraphics[width=\columnwidth]{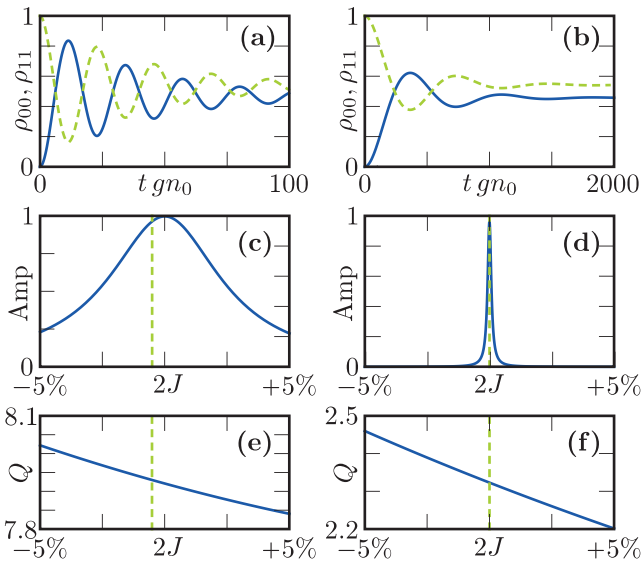}
\end{center}
\caption{Detection of a coherently occupied phonon mode with $|\beta_\pp| = 5$ and frequency $\omega_\pp = 10 \, gn_0$ in one
dimension (left) and two dimensions (right). The detector is tuned close to resonance such that $2J \approx \omega_\pp$. The resulting oscillations
of $\rho_{00}$ (solid blue) and $\rho_{11}$ (dashed green) (a) in one dimension with Q-factor $Q =7.93$ and (b) in two dimensions with
$Q = 2.32$ are clearly visible. Panels (c) and (d) show the sensitivity of the oscillation amplitude (solid blue) on changes of $2J$ away from $\omega_\pp$, realized as $L$ is varied away from its near-resonant value, and $\sigma$ fixed. Panels (e) and (f) show the dependence of
the Q-factor upon the same changes in $L$. The values of $2J$ corresponding to those chosen for the examples (a) and (b) are indicated by the vertical dashed green lines. 
The specific parameters of the Bose gas in 1D are $T = 0.01 \, gn_0/k_B$ and $n_0 \xi = 1$; in 2D we set $T = 0.1 \, gn_0/k_B$, $n_0 \xi^2 = 10$. The system volume ist set to $\Vc = (100 \, \xi )^D$. The parameters of the detector in 1D are
$\kappa = 0.5 \, g$, $\sigma = 0.3 / \abs{\pp}$, $L = 0.62 / \abs{\pp}$ and in 2D we set $\kappa = 2 \, g$, $\sigma = 0.05 / \abs{\pp}$,
$L = 0.14415 / \abs{\pp}$. The impurity mass is $m_a = 0.5  \, m_b$, and the barrier height $\Delta V = D \omega_a$ in both cases.}
\label{coherent examples}
\end{figure}

We next show how to detect the coherent occupation $|\beta_\pp|^2$ of a given mode $\pp$ on top of the thermal fluctuations of the Bose gas.
As shown in \secr{coherent driving}, a coherent density wave induces coherent oscillations of the detector, damped by coupling to the
thermal background. We have already shown that, for these oscillations to be clear, we must have that the Q-factor $Q = \Omega_\pp/2 \bar{k}$ is larger than or on the order of unity, and the detuning $\delta_\pp$ is on the order or smaller than $\Omega_\pp$. We add here an additional requirement from the practicality of observing oscillations within a reasonable time-scale, namely the resonant Rabi frequency $\Omega_\pp$ has to be not much smaller than the natural Bose gas frequency scale $g n_0$, which for realistic experimental parameters $gn_0$ is of the order of 1kHz~\footnote{Values taken from the experiment by Catani {\it et al.}~\cite{Catani-PRA-2012}.}. This last requirement amounts to $p \sigma \lesssim 1$, and thus limits us to wavelengths $2 \pi / p$ on the order of tens of nm or larger.


In \fir{coherent examples} we show examples in one and two dimensions for which it is possible to simultaneously meet all of these requirements and observe oscillations of $\rho_{00}(t)$. In both cases the detector is driven by a coherently occupied high-energy mode with frequency $\omega_\pp = 10~gn_0$, which for $D=1$ ($D=2$) typically corresponds to a wavelength on the order of $\lambda_\pp \sim 100~\text{nm}$ ($\lambda_\pp \sim 1~\mu\text{m}$).


The examples also illustrate the main limitation of our detector as a detector of coherent density waves. For $Q \gtrsim 1$ the width of the resonance is of the order $\Omega_\pp$ and thus the fraction to which $2J$ must be tuned close to $\omega_\pp$ is of the order $\Omega_\pp / \omega_\pp$. For increasingly large $\omega_\pp / g n_0$ this fraction becomes very small, as $\Omega_\pp$ remains on the order of $gn_0$ or less. The situation is worse for higher $D$. The sensitivity with respect to $2J$ is illustrated in Figs.~\ref{coherent examples}(c) and (d) for one and two dimensions, respectively. Conversely, the Q-factor is not so sensitive to the value of $2J$, as shown in Figs.~\ref{coherent examples}(e) and (f), but remains close to unity. Coherent density waves with frequencies $\omega_\pp = 10~gn_0$ are thus at the upper end of what could be realistically detected, with the sensitivities and $Q$-factors at lower frequencies being much improved. 


Finally, note that if one is interested only in whether or not a given mode is occupied, the mere presence of oscillations
yields sufficient information. However, scanning through the resonance should, in principle, yield enough data to determine both the energy of
the mode occupied (from resonant frequencies), the axis along which its momentum is directed (by observing the dependence
of oscillation amplitude on $\LL$), and its occupation $|\beta_\pp|^2$ (from the shape of the resonance and the frequencies
of the oscillations).


\section{Parallel measurements}
\label{prepmes}

It is of course always necessary to perform multiple measurements in order to obtain estimates of the impurity qubit populations that are sufficiently precise for the desired application of the  detector. Realizing the whole measurement procedure (creation of Bose gas and detector, evolution and measurement) a large number of times $M$ results in $M$ independent measurements of the impurity in the energy eigenbasis and thus provides an estimate of the excited state population $\rho_{11}$ with binomial uncertainty $\Delta \rho_{11} = \sqrt{\rho_{11}(1 -\rho_{11}) /M}$ (see~\apr{app:measurement})~\footnote{Similar considerations hold for an estimate of $\rho_{00}$}. However, it is more economical to perform $M$ simultaneous measurements of the states of $M$ different impurities interacting with the same single realization of the Bose gas and use those results to estimate $\rho_{11}$. 

A crucial question is whether in the latter case the state of each impurity is unaffected by and independent from the others, thus ensuring that the precision and accuracy of the combined measurement scales in the same way as for multiple detectors, each interacting with its own separate realization of the Bose gas. Neglecting direct interactions of the detectors by assuming non-overlapping well-functions, we need only investigate the interaction of the detectors via the Bose gas. In the following, we examine this problem separately for the cases of incoherent damping due to thermally occupied Bogoliubov modes, and oscillatory driving due to coherently occupied modes. We will focus on the most promising set-up, namely a linear array of double-well potentials, in combination with a preparation and measurement scheme as described in \apr{app:preparation and measurement}.


\begin{figure}[t]
 \begin{center}
   \includegraphics[width=0.8\columnwidth]{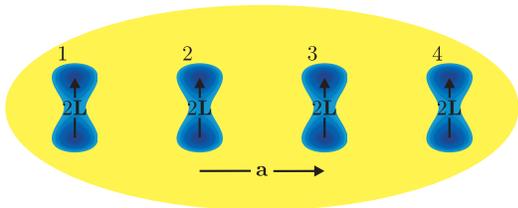}
 \end{center}
 \caption{(Color online) If an array of impurities with lattice vector $\mathbf{a}$ is arranged such that $\mathbf{a}\cdot \LL = 0$ and $|\mathbf{a}|\gg L$, no Bose gas-mediated correlations between the impurities can arise.}
\label{fig:multipleImpurities}
\end{figure}


In the case of incoherent damping, our Born-Markov approach for a single impurity can be extended to describe the dynamical generation of correlations between the impurities. This is done in \apr{app:multipleImpurities}. The impurities affect each other in two distinct ways: a direct, coherent interaction mediated by the exchange of phonons, and an indirect, incoherent growth of classical correlations due to their mutual interaction with the background gas. Both of these effects are strongly dependent on the geometry and the number of spatial dimensions. This ultimately stems from the directional dependence of the phonon radiation emanating from each double-well impurity, which is strongest in the direction parallel to $\mathbf{L}$, but vanishes (at large distances) in the directions perpendicular to $\mathbf{L}$. This is entirely analogous to the directionality of the sound emanating from a classical acoustic dipole.

In two or three dimensions, it is possible to place an array of double-well impurities separated by a lattice vector $\mathbf{a}$ such that $\mathbf{a}\cdot\mathbf{L} = 0$ and $|\mathbf{a}|\gg L$
(see \fir{fig:multipleImpurities}). The generation of correlations between the impurities is completely suppressed in such an arrangement. In any other configuration, both the direct and indirect interactions decay slowly with distance between the impurities, asymptotically as $r^{-(D-1)/2}$. 

In contrast, in one dimension, the impurities are constrained to lie along a line parallel to $\mathbf{L}$; furthermore, both the direct and indirect interactions do not decay with distance. Therefore, the correlations between impurities cannot be neglected in one dimension. In configurations where correlations are important, a more detailed analysis would be required in order to interpret experiments using multiple detectors in a single realisation of a one-dimensional Bose gas. This analysis could in principle be carried out using the microscopic equation of motion presented in \apr{app:multipleImpurities}. Note that the orientation required to minimize correlations, i.e., $\mathbf{a} \perp \LL$ may conflict with some readout schemes, such as the band-mapping TOF image described in~\apr{app:measurement}. In this case $\mathbf{a} \parallel \LL$ is the optimal configuration~\footnote{Simon F{\"o}lling, private communication}. 

Now we consider the case where multiple detectors interact with a Bogoliubov mode in the coherent state $\ket{\beta_{\mathbf{p}}}$. To the order of approximation considered, the mode in question remains in a product state with the impurity detectors. Therefore, to the same order of approximation, the impurities do not become correlated with each other via their mutual interaction with this mode. Nevertheless, the impurities could still conceivably affect each other via their back-action on the state of the gas.

The effect of each impurity on the Bose gas is to generate additional phonons. If the number of extra phonons is small compared to the mean number of phonons already present, then the back-action of the detectors can be neglected. In \apr{app:back-action} we derive the conditions for this to be the case. We find that, in the majority of configurations, the displacements enacted by multiple impurities do not add up constructively. Therefore, the effect of several impurities on a coherent state of the Bose gas is no greater than the effect of one. Taking our specific example of a linear array of $M$ impurities, we find that the back-action of the detectors is negligible unless $2\pi( n - 1/M) \lesssim \mathbf{p} \cdot \mathbf{a} \lesssim 2\pi( n + 1/M)$ for some integer $n$, a situation that is easy to avoid in practice by choosing the lattice vector $\mathbf{a}$ appropriately.

It follows then that using an array of many double-wells, where numbers in the thousands are easily achievable, we can obtain accuracies corresponding to a large number of measurements in the time it takes to perform a single measurement, and, in principle, without destroying the state of the Bose gas being measured.


\section{Conclusion}
\label{sec:conclusion}

Quantum probes in general and localized atomic impurities in particular offer an alternative way of probing many-body quantum systems,
with the potential to be non-destructive. Specifically, probing phononic exciations is important, for example, to measure temperature, and in the context of the simulation of gravitational models and shock-wave formation. Here we have shown how to make use of the versatility of the double-well qubit to probe the phononic excitations of a weakly-interacting cold Bose gas. We have shown that, due to the ability to localize the probe to tens of nanometers or tune its resonant energy down to the nano-Kelvin regime, it provides a flexible alternative to other probing techniques, for example, imaging with light. Our technique has the additional benefit of being potentially {\it non-destructive}. In an experiment, a suitably aligned array of $M \sim 1000$ double wells allows the user to realize many independent detectors simultaneously.

A major part of our proposal involves monitoring the non-equilibrium dynamics of the impurity in order to infer properties of the Bose gas. This approach relies on, and is thus only as accurate as, the Bogoliubov description of the Bose gas. More generally, deviations of the detector from the expected behaviour may be useful for exploring the limits of the Bogoliubov description. 

We have focused on the use of the impurity as a probe that measures the occupation of certain phonon modes. If the occupation of these modes is known by other means, the probe could also measure the density of states of the Bose gas at a specific energy. We also note that this same system may be used to absorb and thereby remove specific excitations from a phonon environment.

\begin{acknowledgements}

DH thanks Simon F{\"o}lling  for useful discussions, and acknowledges support from the Studienstiftung des deutschen Volkes and Erasmus Placements. MTM acknowledges financial support from the UK EPSRC. THJ and DJ thank the National Research Foundation and the Ministry of Education of Singapore for support. MB and MBP acknowledge financial support from an Alexander von Humboldt Professorship, the EU Integrating project SIQS and the EU STREP EQUAM. We gratefully acknowledge financial support from the Oxford Martin School Programme on Bio-Inspired Quantum Technologies.

\end{acknowledgements}

\bibliography{Lit}

\clearpage

\begin{appendix}

\section{Experimental realization of the detector}
\label{app:experimental realization}

A one-dimensional array of double wells can be created by an optical lattice potential $V(\rr)$
which is obtained by superposing two pairs of laser beams with wave vectors $\mathbf{k}$ parallel
to $\mathbf{L}$ and wavelengths $\lambda$ and $\lambda/2$, respectively~\cite{Sebby-Strabley-PRL-2007,Foelling-Nature-2007}.
Introducing the axial coordinate $r_{\parallel}$ in the direction of $\mathbf{k}$ and the radial
coordinate $r_{\perp}$ perpendicular to $\mathbf{k}$ we can write the potential as
\begin{equation*}
\begin{split}
	V (\rr) &= V_0\left[\cos^2 \left(\frac{2 \pi r_{\parallel}}{\lambda}\right)
	- 4 v \cos^2 \left(\frac{\pi r_{\parallel}}{\lambda}\right)\right]\\
	&\times \exp\left(- \frac{1}{2}\frac{r_{\perp}^2}{w^2}\right),
\end{split}	
\end{equation*}
where $w$ is the width of the beams.
The well separation $2L$, curvature $\omega_a$, and sizes of the barriers inside and between the double wells are determined by the relative intensity $4v$ of the two lattices. Specifically, as $v$ goes from $0$ to $1$, the well separation $2L$ goes from $\lambda/2$ to $\lambda \arccos(v)/2\pi \approx \lambda/2 \, (1 - 2 v/\pi) $, curvature $\omega_a$ from $(2\pi/\lambda) \sqrt{2 V_0 /m_a} $ to $(2\pi/\lambda) \sqrt{ 2 V_0 (1-v^2)/m_a}$, double-well barrier height $\Delta V$ from $V_0$ to $V_0 \,(1 - v)^2$, and inter-double-well barrier height from $V_0$ to $V_0 \,(1+ v)^2$. Choosing $v \lesssim 1$ therefore results in an array of separated double-wells, each with $L$, $\omega_a$ and $\Delta V$ on the order of $\lambda$, $\sqrt{2 V_0 /m_a\lambda^2}$ and $\omega_a$ respectively.

It follows then that using typical experimental parameters $V_0 \lesssim 10^3\, E_{r}$, $m_a \approx 10^2 \, \text{amu}$ and $w, \lambda \approx \mu \text{m}$, with $E_r = 4 \pi^2 / 2 m_a \lambda^2$ the recoil energy, it is possible to achieve spatial scales as short as $\sigma \approx 10 \, \mathrm{nm}$ and energy scales as high as $\omega_a/2\pi \approx 100 \, \text{kHz}$, respectively~\footnote{Explicitly, using the parameters in the experiment of Fukuhara {\it et al.}~\cite{Fukuhara-PRA-2009}, $\lambda = 532\,\text{nm}$, $m_a = 174 u$ and $V_0 = 20 \, E_r$, the impurity width is $\sigma \approx 40 \, \text{nm}$, the trapping frequency $\omega_a/2 \pi \approx 40 \, \text{kHz}$.
However, one can further increase the lattice depth to $\sim 100 \, E_r$ to achieve the given value $\sigma \approx 20 \, \text{nm}$, while $\omega_a / 2\pi \approx 80 \, \text{kHz}$.}. 
At the opposite end of the spectrum, using a shallow lattice depth $V_0 \approx 1 \,E_r$, $m_a \approx 10 \, \text{amu}$ and $w, \lambda \approx 10 \mu \text{m}$, one may achieve spatial scales $\sigma \approx 1 \, \mu \text{m}$ and energy scales $\omega_a/2\pi \approx 1 \, \text{kHz}$.
It is also possible to simultaneously realize $L \ll \lambda, w$ and thus $\Delta V / \omega_a \approx 1$ while exploring a broad range of $L / \sigma$ for which $2J$ is smaller than or the same order as $\omega_a$.

Fluctuations in the individual and parallel intensity and phase of the lasers creating the two lattices cause the dephasing of the impurity. For example, in Ref.\ \cite{Foelling-Diss} it is reported that frequency fluctuations between the lattice lasers amounting to $< 100$ kHz result in fluctuations of the bias between the wells $\abs{V_a(\LL) - V_a( - \LL)} < 100$ Hz. However, the total magnitude of such fluctuations can be kept small such that coherence times of the atoms in the double wells of more than 10 ms have already been achieved~\cite{Anderlini-Nature-2007, Foelling-Nature-2007}.


\section{Spectral densities for general dispersion}
\label{app:spectral density}

We evaluate the general form of the spectral density in $D$
dimensions [cf.~\eqr{sdens}] using the full Bogoliubov dispersion
relation $\omega_\qq = \sqrt{E_\qq (E_\qq + 2 g n_0 )}$, which enters in the decay rates given in
Eqs.~\eqref{dampingrates}, in the continuum limit. In this limit $\Vc^{-1} \sum_\qq \rightarrow (2\pi)^{-D} \int \dd \Omega_D \int \dd q q^{D-1}$, 
where $\Omega_D$ is the surface of the unit sphere in $D$ dimensions, we find 
\begin{align*}
  \begin{split}
    \Jc_D (\omega)  = & \kappa^2 \frac{\Vc}{(2\pi)^D} \int \dd q \delta(\omega-\omega_\qq) q^{D-1} \int d\Omega_D \abs{M_\qq}^2 \\
    = &\kappa^2 \frac{n_0}{(2\pi)^D} \int \dd \omega_\qq \delta( \omega - \omega_\qq )  \frac{ E_{q(\omega_\qq)} }{ \omega_\qq } \left( \frac{ \partial \omega_\qq }{ \partial q } \right)^{-1} \\
    & \, \, \times e^{-q(\omega_\qq)^2\sigma^2/2} \int \dd \Omega_D \sin^2(\qq\cdot \LL) \, ,
  \end{split}
\end{align*}
Here $q(\omega_\qq)$ is the inverse of the dispersion relation $\omega_\qq$ and the angular integrals read 
\begin{equation*}
\int \dd \Omega_D \sin^2(\qq\cdot \LL) = \left\{ 
\begin{array}{ll} 2 \sin^2(q L) & D = 1 \\
\pi [1 - \Bc_0(2 q L) ] & D = 2\\
2\pi [1 - \text{sinc}(2 q L ) ] & D = 3 
\end{array} 
\right \} \, . 
\end{equation*}
Explicitly inserting the expressions for the inverse and the Jacobian of the Bogoliubov dispersion, 
 \begin{align*}
\xi^2 q(\omega)^2 & = 2 \left( 1 + \sqrt{1 + (\omega/gn_0)^2} \right)\, ,\\ 
(\xi g n_0)^{-1}  \frac{ \partial \omega_\qq}{\partial q} & =  \xi q \left( \frac{E_\qq + g n_0 }{ \omega_\qq} \right) ,
\end{align*}
the spectral densities read 
\begin{align*}
\frac{\Jc_1(\omega)}{gn_0}  = & \frac{(\kappa/g)^2}{\pi n_0 \xi}  \frac{1}{q(\omega) \xi} \frac{E_{q(\omega)}}{E_{q(\omega)} + 2 g n_0} e^{-\half q(\omega)^2 \sigma^2} \sin^2\left(q(\omega) L\right) \, , \nonumber \\
\frac{\Jc_2(\omega)}{gn_0}  = & \frac{(\kappa/g)^2}{2 \pi n_0 \xi^2}  \frac{E_{q(\omega)}}{E_{q(\omega)} + 2 g n_0}  \,  e^{-\half q(\omega)^2 \sigma^2} \left[1 - \Bc_0(2 q(\omega) L) \right]\, ,\nonumber\\
\frac{\Jc_3(\omega)}{gn_0}  = & \frac{(\kappa/g)^2}{2 \pi^2 n_0 \xi^3}  \, q(\omega) \xi \,  \frac{E_{q(\omega)}}{E_{q(\omega)} + 2 g n_0}  \\
& \, \, \times  e^{-\half q(\omega)^2 \sigma^2} \left[1 - \text{sinc}(2 q(\omega) L) \right]. \nonumber
\end{align*}


\section{Back-action of impurities in the mean-field approximation}
\label{app:back-action}

Here we examine the conditions under which we can neglect the back-action of impurities on a Bogoliubov mode in the coherent state $\ket{\beta_{\mathbf{p}}}$. Within the mean-field approximation introduced in \secr{coherent driving}, the presence of the impurity causes a displacement of the Bogoliubov coherent state in phase space. The displacement operation is defined by $\hat{D}(\alpha_{\mathbf{p}})\ket{\mathrm{vac}} = \ket{\alpha_{\mathbf{p}}}$, i.e.\ a displacement creates the coherent state $\ket{\alpha_{\mathbf{p}}}$ from the Bogoliubov ground state $\ket{\mathrm{vac}}$, which satisfies $\an{b}_{\mathbf{q}}\ket{\mathrm{vac}} = 0$ for all $\mathbf{q}$. If the magnitude of the displacement parameter $\alpha_{\mathbf{p}}$ is much smaller than that of the coherent state amplitude $\beta_{\mathbf{p}}$, then we can say that the back-action of the impurities is negligible. In other words, we would like to show that the impurities do not create enough additional phonons in the gas to have an appreciable effect on each other's evolution.

If we consider $M$ double-well impurities centered at the positions $\mathbf{x}_{i}$, the corresponding time-dependent displacement parameter is given by 
\begin{align*}
\alpha_{\mathbf{p}}(t) = -\ii\kappa S_{\mathbf{p}} \ee^{-\sigma^2 p^2/4}\sum_{i=1}^{M}&\ee^{\ii\mathbf{p}\cdot \mathbf{x}_{i}}\int_0^{t}\dd s\,  \ee^{\ii\omega_{\mathbf{p}} s}[ \cos(\mathbf{p}\cdot \mathbf{L}) \notag \\ & - \ii\sin (\mathbf{p}\cdot \mathbf{L}) \av{\sigx^{(i)} (s)}],
\end{align*}
where the operator $\sigx^{(i)}$ pertains to impurity $i$. For simplicity, we take a constant $\av{\sigx^{(i)} (s)} = 1$, which amounts to the assumption that each impurity is static and localised in the left minimum of its double-well potential. In the majority of situations this assumption represents an over-estimate of $\alpha_{\mathbf{p}}(t)$, since the oscillations of the impurity dynamical variables are assumed to be far detuned from resonance with $\omega_{\mathbf{p}}$, and are generally out of phase with each other. We further assume, for the sake of concreteness, that the detectors are arranged in a 1D lattice with lattice parameter $\mathbf{a}$. We thus obtain
\begin{equation}
\label{alphaAssumptions}
|\alpha_{\mathbf{p}}(t)|^2 = \frac{\kappa^2 S_\pp^2}{\omega^2_\pp}\frac{\sin^2 (M\mathbf{p}\cdot \mathbf{a}/2)}{\sin^2 (\mathbf{p}\cdot \mathbf{a}/2)}\sin^2(\omega_\pp t/2) \ee^{-\sigma^2p^2/2} .
\end{equation}

Neglecting the oscillatory time-dependence, the contribution of each individual impurity to \eqr{alphaAssumptions} is of order $ \kappa^2 S^2_\pp\ee^{-\sigma^2p^2/2}/\omega^2_\pp$.  Since $S^2_{\mathbf{p}}\ee^{-\sigma^2 p^2/2} \sim |M_{\mathbf{p},01}|^2$, we find that the displacement due to a single impurity is small compared to $\beta_{\mathbf{p}}$ if
\begin{equation}
\label{amplitudeCondition}
\frac{\Omega_{\mathbf{p}}}{\omega_{\mathbf{p}}}\ll|\beta_{\mathbf{p}}|^2  ,
\end{equation}
where the resonant Rabi frequency is $\Omega_{\mathbf{p}} = 2 \kappa | M_{\pp,10}^* \beta_\pp |$. \eqr{amplitudeCondition} is easily satisfied within the regime of validity of the rotating wave approximation, where $\Omega_{\mathbf{p}}/\omega_{\mathbf{p}}\ll 1$. 

With $M$ impurities we also have to include the factor $\sin^2 ( M\pp \cdot \mathbf{a} /2) / \sin^2 (\pp \cdot \mathbf{a} /2)$, which is typically of order unity and only takes significant values ($\sim M^2$) when $\qq \cdot \mathbf{a}$ differs from an integer multiple of $2 \pi$ by an amount of the order $2 \pi /M$ or less. Hence, for nearly all cases the evolution of the impurities is unaffected by the excitations they themselves create in the Bose gas, since their contributions are generally out of phase and therefore do not add up significantly.


\section{General solution of the Bloch equations}
\label{app: diagonalization}

We solve the Bloch equations~\eqref{bloch} for the fully coherent case ($\bar{k} = 0$) to obtain the Rabi oscillations described by
Eq.~\eqref{pdynamicscoherent}, and for the resonant case $\delta_\pp=0$ with damping ($\bar{k} \neq 0$), which is used in
Sec.~\ref{coherent detection}. In both cases we are interested in the population $\rho_{00}(t)$ with the initial condition
$\rho_{00}(0) = 1$. The population $\rho_{00}$ is related to the Bloch vector $\vv$ through $\rho_{00} = (1-v_z)/2$.

The Bloch equations in matrix form read $\dd\vv/\dd t = M\vv + \mathbf{u}$ with $\mathbf{u} = (0,0,k_{+}-k_{-})^{T}$ and
\begin{equation*}
M = \left(
\begin{array}{ccc}
 -\bar{k} & -\delta_{\pp}  & 0 \\
 \delta_{\pp}  & -\bar{k} & -\Omega_{\pp}  \\
 0 & \Omega_{\pp}  & -2\bar{k} \\
\end{array}
\right)\,.
\end{equation*}
In the coherent case $\bar{k}=0$ the diagonal elements of $M$ are zero and $\mathbf{u} = 0$. The Bloch vector can be
found directly from $\vv(t) = \ee^{M t}\vv(0)$ with the initial condition $\vv(0) = (0,0,-1)^{T}$, for which we obtain
\begin{equation*}
\begin{split}
		v_x(t) &= \frac{\delta_{\pp}\Omega_{\pp}}{\tilde{\Omega}_{\pp}^2}[\cos (\tilde{\Omega}_{\pp} t)-1]\,,\\
		v_y(t) &= \frac{\Omega_{\pp}}{\tilde{\Omega}_{\pp}}\sin(\tilde{\Omega}_{\pp} t)\,,\\
		v_z(t) &= -\frac{\delta_{\pp}^2}{\tilde{\Omega}_{\pp}^2}-\frac{\Omega_{\pp}^2 }{\tilde{\Omega}_{\pp}^2}\cos (\tilde{\Omega}_{\pp} t)\,.
\end{split}
\end{equation*}
Equation~\eqref{pdynamicscoherent} then follows from the solution for $v_z(t)$.

In the case with damping but no detuning the solution is of the form $\vv(t) = \tilde{\vv}(t) + \vv^{\infty}$, where $\tilde{\vv}(t)$
solves $\dd\tilde{\vv}/\dd t = M\tilde{\vv}$ and $\vv^{\infty}$ is the stationary state in the long-time limit. The latter
is found from $\vv^{\infty} = -M^{-1}\mathbf{u}$ and given by $v_x^{\infty}=0$,
\begin{equation*}
\begin{split}
		v_y^{\infty} &= \Omega_{\pp}\,\frac{k_{-}-k_{+}}{2\bar{k}^2+\Omega_{\pp}^2}\,,\\
		v_z^{\infty} &= \bar{k}\,\frac{k_{+}-k_{-}}{2\bar{k}^2+\Omega_{\pp}^2}\,.
\end{split}
\end{equation*}
The solution for $\tilde{\vv}(t)$ can be expressed in terms of the eigenvectors $\mathbf{m}_j$
and eigenvalues $\lambda_j$ of the matrix $M$ as
\begin{equation*}
	\tilde{\vv}(t) = \sum_{j=1}^3 c_j\mathbf{m}_j\ee^{\lambda_j}\,,
\end{equation*}
where the coefficients $c_j$ are fixed by the initial conditions
$\tilde{\vv}(0) = \vv(0)- \vv^{\infty} = (0,-v_y^{\infty},-v_z^{\infty}-1)$. This procedure finally
yields the relevant component of the Bloch vector
\begin{equation*}
\begin{split}
	\tilde{v}_z(t) &= \frac{2 \Omega_{\pp}v_y^{\infty} - (v_z^{\infty}+1)(\bar{k}+w_{\pp})}{2 w_{\pp}}\, \ee^{-\frac{1}{2}(3\bar{k}+w_{\pp})t} \\
	& -\frac{2 \Omega_{\pp}v_y^{\infty} - (v_z^{\infty}+1)(\bar{k}-w_{\pp})}{2 w_{\pp}}\, \ee^{-\frac{1}{2}(3\bar{k}-w_{\pp})t}\,.
\end{split}	
\end{equation*}
The quantity $w_{\pp} = (\bar{k}^2 - 4\Omega_{\pp}^2)^{1/2}$ is real-valued for strong damping
$\bar{k}\gg \Omega_{\pp}$ and purely imaginary in the strong driving regime $\Omega_{\pp}\gg\bar{k}$,
in which case the population $\rho_{00}(t) = [1 - v_z^{\infty} - \tilde{v}_z(t)]/2$ exhibits clear
oscillations.


\section{Uncertainties in the temperature measurement}
\label{sec:Terror}

Now we quantify the uncertainty incurred in an experiment to determine the Bose gas temperature resulting from the finite number of measurements. We do this both for the equilibrium method and the non-equilibrium method. 

\setcounter{paragraph}{0}
\paragraph{Equilibrium method.}

When extracting the temperature $T_0$ from the equilibrium population $\rho_{11} = k_+ /(k_+ + k_-)$, given knowledge of $2J$, the value of $T_0$ responds to small changes $\delta \rho_{11}$ of the equilibrium population as
\begin{equation*}
\frac{\delta T_0}{T_0} = \frac{\rho_{11}}{T_0} \left( \frac{\partial \rho_{11}}{\partial T_0} \right)^{-1} \frac{ \delta \rho_{11} }{\rho_{11}} \, .
\end{equation*}

Assuming that the estimate of $\rho_{11}$ is obtained by $M$ independent measurements of the state of the qubit in the energy basis, we find that the expected uncertainty $\Delta T_0$ of the temperature $T_0$ is 
\begin{equation*}
  \frac{\Delta T_0}{T_0} =  \frac{1}{ \sqrt{M} }\frac{k_B T_0}{2J} \frac{(2 n(2J,T_0) + 1)(n(2J,T_0) + 1)^{1/2}}{ \ee^{2J/k_B T_0}  [n(2J,T_0)]^{3/2}} \, ,
\end{equation*}
where we have used the variance $p(1-p)$ of a binomial variable that takes $0$ and $1$ with probability $(1-p)$ and $p$, respectively, and the $1/\sqrt{M}$ scaling of the standard deviation of an average over $M$ measurement outcomes.

\paragraph{Non-equilibrium method.} 

If there is a small change $\delta \bar{k}(T_0)$ in the rate $\bar{k}(T_0)$ the corresponding change $\delta T_0$ in our estimate of $T_0$ from \eqr{decay} is given by
\begin{equation*}
  \frac{ \delta T_0}{T_0} = \frac{k_B T_0}{2J}  \frac{2 n(2J,T_0) +1}{2 \ee^{2J /k_B T_0} n^2(2J,T_0)} \frac{ \delta \bar{k}(T_0)}{\bar{k}(T_0) } .
\end{equation*}

How accurately the rate $\bar{k}(T_0)$ can be measured is in turn fundamentally limited by the accuracy to which $\rho_{11}$ can be measured. As an example, the rate 
\begin{equation*}
	 \bar{k} (T_0) =  \frac{1}{2t_0} \ln \left[ \frac{1-\rho_{11} (\infty)}{\rho_{11} (t_0)-\rho_{11} (\infty)} \right],
\end{equation*}
can be determined from the values of $\rho_{11}(t)$ at two times, firstly a time in the middle of the decay, e.g., $t_0 = \ln (2) / 2 \bar{k}(T_0)$, and second a time at the end of the decay $t_1 \gg 1 / \bar{k}(T_0)$, effectively $t_1 = \infty$. For small changes $\delta \rho_{11} (t_0)$ and $\delta \rho_{11} (\infty)$ in these values, the small change $\delta \bar{k}(T_0)$ in rate $\bar{k}(T_0)$ is thus given by
\begin{equation*}
	 \frac{\delta \bar{k}(T_0)}{\bar{k}(T_0)}  = \frac{ \delta \rho_{11} (t_0) - \delta \rho_{11} (\infty)}{2 \ln (2) [ 1- \rho_{11} (\infty)]} \, , 
\end{equation*}
where the denominator is roughly on the order or unity.
Assuming independent measurements of $\rho_{11}(t)$ at different times and that the estimate of $\rho_{11}(t)$ is obtained as above ($M$ independent measurements) we obtain the expected uncertainty
\begin{align*}
  \begin{split}
    \frac{ \Delta T_0}{T_0} = &  \frac{k_B T_0}{2J}   \frac{2 n(2J,T_0) +1}{2 \ee^{2J /k_B T_0} n^2(2J,T_0)} \\
    & \times \frac{\sqrt{ \rho_{11} (t_0)[1-\rho_{11} (t_0)] + \rho_{11} (\infty)[1-\rho_{11} (\infty)]}}{\sqrt{M}2 \ln (2) [ 1- \rho_{11} (\infty)]}  \, ,
  \end{split}
\end{align*}
for the estimate of the temperature $T_0$. 

Assuming for example that the impurity parameters are chosen such that $\rho_{11} (t_0), \rho_{11} (\infty) \approx 1/2$, the resulting uncertainty obeys
\begin{align*}
  \frac{ \Delta T_0}{T_0} \approx \frac{1}{\sqrt{M}} \frac{k_B T_0}{2 J}   \frac{2 n(2J,T_0) +1}{2 \sqrt{2} \ln (2) \ee^{2J /k_B T_0} n^2(2J,T_0)} .
\end{align*}


\section{Preparation and measurement of a single qubit}
\label{app:preparation and measurement}

The measurement schemes proposed in \secr{sec:application} require the controlled preparation of the double-well qubit in its two lowest eigenstates $\ket{0}$ or $\ket{1}$, as well as the measurement of the qubit in its eigenbasis. Here, we discuss the experimental techniques suitable to realize these requirements.

\subsection{Preparation of the initial qubit state}
\label{app:preparation}

For all of the procedures we propose the qubit is prepared in either of the two states $\ket{0}$ or $\ket{1}$. We now briefly discuss, how the initial state can be prepared experimentally. 

We consider an array of double wells formed by a superposition of two lattices with wavelengths $2 \lambda$ and $\lambda$ as discussed in \apr{app:experimental realization}. In such a setup the symmetric state is prepared as follows: first, the $2\lambda$-lattice is unit-loaded with impurity atoms by slowly increasing its intensity. Second, the intensity of the $\lambda$-lattice is increased quickly to the desired value, while avoiding vibrational excitation \cite{Sebby-Strabley-PRL-2007}. The barrier acts like a beam splitter; the impurity atoms are now in the symmetric state. With a $\pi/2$-phase kick the impurity can then be lifted to the excited state $\ket{1}$. This may be realized by switching back to the $2\lambda$-lattice, exciting the impurity, and then raising the $\lambda$-lattice again. 

\subsection{Measurement of a single qubit}
\label{app:measurement}

All of the applications introduced in the previous section reduce to
measuring the population of the energy eigenstate $\rho_{00}$ (equivalently $\rho_{11}$). Moreover, in \secr{temperature detection} we assumed this was done via repeated measurement in the energy eigenbasis. 

This is indeed possible with a cold atom setup. For example, one appropriate
technique is a band-mapping variant of a time-of-flight
(TOF) absorption measurement~\cite{Foelling-Diss,Kastberg-PRL-1995,Greiner-PRL-2001,Sebby-Strabley-PRL-2007}. In the above setup the $\lambda$ lattice is
slowly lowered, adiabatically mapping the $\ket{0}$ and $\ket{1}$ states into the first and second Brillouin zones (bands), respectively, of the $2\lambda$ lattice. In a second step, the $2\lambda$ lattice is
slowly lowered, adiabatically mapping the quasi-momentum states to real momentum states, thus mapping $\ket{0}$ and $\ket{1}$ to non-overlapping regions of momentum space. Therefore a TOF image of the resulting expansion measures an impurity in the energy eigenbasis.

Repeating this procedure for $M$ identical copies of the whole system, the fraction of measurements obtaining the impurity ground state provides an estimate of $\rho_{00}$ with binomial uncertainty $\Delta \rho_{00} = \sqrt{\rho_{00}(1 -\rho_{00}) /M}$.


\section{Master equation for multiple impurities}
\label{app:multipleImpurities}

We consider a system of $M$ impurity detectors, each trapped in separate but otherwise identical double-well potentials. We assume, partly for simplicity and partly as it is experimentally likely, that all the detectors are aligned parallel to each other, so that the two minima of the potential confining impurity $i$ are located at the positions $\mathbf{x}_{i} \pm \mathbf{L}$.  Neglecting a constant, the Hamiltonian of the impurities is 
\begin{equation*}
\hat{H}_a = \sum\limits_{i=1}^{M} J\sigz^{(i)}.
\end{equation*}
Making the same assumptions described in \secr{sec:model}, we find that the interaction of the impurities with the gas is described by the Hamiltonian
\begin{align*}
\hat{H}_{ab} = & \sum\limits_{i=1}^{M} \sum\limits_{\mathbf{q}} \left ( M_{\mathbf{q},01} e^{\ii \mathbf{q}\cdot \mathbf{x}_i} \cre{b}_{\qq} + M_{\mathbf{q},01}^{\ast} e^{-\ii \mathbf{q}\cdot \mathbf{x}_i} \an{b}_{\qq}\right ) \otimes \sigx \notag \\ & + \sum\limits_{i=1}^{M} \sum\limits_{\mathbf{q}}  \left ( \bar{M}_{\mathbf{q}} e^{\ii \mathbf{q}\cdot \mathbf{x}_i} \cre{b}_{\qq} +  \bar{M}^{\ast}_{\mathbf{q}} e^{-\ii \mathbf{q}\cdot \mathbf{x}_i} \an{b}_{\qq} \right ) \otimes \id,
\end{align*}
where the matrix elements $M_{\qq,01}$ and $\bar{M}_{\qq} = \half(M_{\qq,00} + M_{\qq,11})$ are defined by \eqr{simpel}. 

Under the same approximations as described in \secr{damping}, we derive a Lindblad master equation describing the impurity dynamics induced by their interaction with the thermally occupied Bogoliubov modes. This takes the form
\begin{align*}
\frac{\partial \an{\rho}_a}{\partial t} = \sum_i \mathcal{L}_{i}[\an{\rho}_a(t)] + \sum\limits_{i\neq j} \mathcal{C}_{ij}[\an{\rho}_a(t)],
 \end{align*} 
where $\mathcal{L}_{i}$ describes the independent evolution of impurity $i$, while $\mathcal{C}_{ij}$ describes the dynamics of correlations between impurities $i$ and $j$ due to their mutual interaction with the spatially correlated reservoir. Explicitly, the local terms read
\begin{align*}
\mathcal{L}_{i}[\an{\rho}] =  & -\ii J  [\sigz^{(i)},\an{\rho} ] \notag \\ & + \sum_{s=\pm} k_{s}\Big(\an{\sigma}^{(i)}_s\an{\rho}(\hat{\sigma}^{(i)}_s)^{\dagger}  -\frac{1}{2}\{(\hat{\sigma}^{(i)}_s)^{\dagger}\an{\sigma}^{(i)}_s,\hat{\rho}\}\Big) ,
\end{align*}
where the local gain and decay rates $k_{\pm}$ are defined by \eqr{dampingrates}. The correlation terms read
\begin{align*}
\mathcal{C}_{ij}[\an{\rho}] = & -\ii \eta^{(ij)}  [\sigp^{(i)}\sigm^{(j)},\an{\rho} ]\notag \\ & +\sum_{s=\pm} \Gamma^{(ij)}_{s}\left (\an{\sigma}_s^{(j)} \an{\rho} (\an{\sigma}_s^{(i)})^{\dagger} - \frac{1}{2} \lbrace (\an{\sigma}_s^{(i)})^{\dagger}\an{\sigma}_s^{(j)},\an{\rho}  \rbrace \right  ) .
\end{align*}
The exchange of phonons leads to coherent coupling of strength $\eta^{(ij)}$ between impurities $i$ and $j$, while the rate of growth of classical correlations between these impurities is controlled by the quantities $\Gamma^{(ij)}_{\pm}$. These parameters can be expressed in terms of the interdependence function
\begin{equation}
\label{intFunction}
\mathcal{F}_{D}(\omega,\mathbf{r}) = \kappa^2\sum_{\mathbf{q}} |M_{\qq,01}|^2 \cos (\mathbf{q}\cdot \mathbf{r}) \delta(\omega - \omega_\qq).
\end{equation}
Specifically, we have
\begin{equation*}
\eta^{(ij)} = \mathcal{P} \int\limits_{0}^{\infty} \dd\omega\;\frac{2\omega}{4J^2 - \omega^2} \mathcal{F}_{D}(\omega,\mathbf{r}_{ij}),
\end{equation*}
where $\mathcal{P}$ denotes the Cauchy principal value and $\mathbf{r}_{ij} = \mathbf{x}_i - \mathbf{x}_j$, while
\begin{align*}
\Gamma_-^{(ij)} & = 2\pi  [n(2J)+1]  \mathcal{F}_D(2J,\mathbf{r}_{ij})\notag \\ 
\Gamma_+^{(ij)} & = 2\pi  n(2J) \mathcal{F}_{D}(2J,\mathbf{r}_{ij}).
\end{align*}

The three parameters $\eta^{(ij)}$, $\Gamma_{-}^{(ij)}$ and $\Gamma_{+}^{(ij)}$ control the rate at which correlations between the impurity detectors are generated throughout the detection procedure. In order to actually evaluate these quantities, one must first carry out $\qq$-summation entering the interdependence function \eqref{intFunction}. The result of this computation is highly dependent on both the number of spatial dimensions and the geometrical configuration of the detectors. In order to understand this dependence, it suffices to consider the function
\begin{equation*}
G_D(q,\mathbf{r}) = \int\limits\dd\Omega_D \sin^2(\qq\cdot \mathbf{L})\cos (\mathbf{q}\cdot \mathbf{r}),
\end{equation*}
which is proportional to the interdependence function \eqref{intFunction} in the continuum limit. The integral extends over the solid angle $\Omega_D$ subtended on the $(D-1)$-sphere by the momentum vector $\qq$. Performing this integral yields
\begin{align}
\label{gFactor1}
G_1(q,r) & = 2\cos(qr) \sin^2(qL) \, , \\ 
\label{gFactor2}
G_2(q,\mathbf{r}) & \approx 2\pi \,\mathcal{B}_0(qr) \sin^2(q \hat{\mathbf{r}}\cdot \mathbf{L})  \, ,\\
\label{gFactor3}
G_3(q,\mathbf{r}) & \approx 4\pi\,\mathrm{sinc}(qr)\sin^2(q \hat{\mathbf{r}}\cdot \mathbf{L}) \, ,
\end{align}
where $\hat{\mathbf{r}} = \mathbf{r}/r$. Our expression for one dimension is exact, however we have employed some approximations in two and three dimensions in order to obtain more intelligible equations. In both two and three dimensions we have made the natural assumption that the impurities are placed far apart, so that $r\gg L$. In two dimensions we have also made the more stringent assumption that $qr\gg 1$ over the frequency range of interest. 

By examining \eqr{gFactor1}, we see that the growth of correlations is unavoidable in one dimension. This is due to the fact, peculiar to one dimension, that density waves generated by each impurity propagate with constant intensity at arbitrarily large distances from the source. In two and three dimensions, on the other hand, energy conservation dictates that the flux of density waves is attenuated over distance, leading to an asymptotic decay of $G_D(q,\mathbf{r})\sim r^{-(D-1)/2}$. Most importantly, the factor $\sin^2(q \hat{\mathbf{r}}\cdot \mathbf{L})$ appearing in Eqs.~\eqref{gFactor2} and \eqref{gFactor3} indicates that \textit{no correlations} are generated for ``perpendicular" configurations, when $\mathbf{r}_{ij}\cdot \mathbf{L} = 0$ for all impurity pairs. This is because of the angular dependence of the phonon radiation emanating from each detector, which at large distances vanishes in the directions perpendicular to $\mathbf{L}$. Likewise, each impurity does not respond to density waves impinging from directions perpendicular to $\mathbf{L}$, as demonstrated in \fir{oscillation amplitude}. Therefore, in two and three dimensions, it is possible to create a one-dimensional array of impurity detectors that can probe a single realisation of a Bose-Einstein condensate without affecting each other's measurements.

\end{appendix}

\end{document}